\documentclass[journal]{IEEEtran}

\usepackage[T1]{fontenc}
\RequirePackage{fix-cm} 

\usepackage{amsmath,amssymb,amsfonts} 
\usepackage{mathtools} 
\usepackage{bm} 
\usepackage{siunitx} 
\usepackage{adjustbox}
\usepackage{amsmath} 
\usepackage{optidef}

\usepackage{graphicx} 
\usepackage{caption} 
\usepackage{subcaption} 
\usepackage{booktabs} 

\usepackage{algorithm}
\usepackage{algorithmic} 

\usepackage{cite} 
\usepackage{url} 
\usepackage{hyperref}
\usepackage[hyphenbreaks]{breakurl}
\hypersetup{colorlinks=true, linkcolor=blue, citecolor=blue, urlcolor=black}

\usepackage{float} 
\usepackage{stfloats} 
\usepackage{afterpage} 
\usepackage{array} 
\usepackage{tcolorbox} 
\tcbuselibrary{skins} 
\usepackage{enumitem} 
\usepackage{amsmath} 
\newtcolorbox{taskdescription}[1][]{
	enhanced,
	boxrule=0.8pt,
	colback=white,
	colframe=blue!50!black,
	arc=4pt,
	boxsep=5pt,
	left=6pt,
	right=6pt,
	top=6pt,
	bottom=6pt,
	fontupper=\small,
	title={\textbf{Task Description for Cooperative Caching Strategy}},
	#1
}

\begin{document}

	\title{LLM-Empowered Cooperative Content Caching in Vehicular Fog Caching-Assisted Platoon Networks} 
\author{Bowen Tan,
	Qiong Wu, \textit{Senior Member, IEEE},
	Pingyi Fan, \textit{Senior Member, IEEE},\\
\hspace{0.5em}Kezhi Wang, \textit{Senior Member, IEEE},
	Nan Cheng, \textit{Senior Member, IEEE},
and	Wen Chen, \textit{Senior Member, IEEE}%
\thanks{This work was supported in part by Jiangxi Province Science and Technology Development Programme under Grant 20242BCC32016, in part by the National Natural Science Foundation of China under Grant 61701197, 62531015, and U25A20399, in part by the Basic Research Program of Jiangsu under Grant BK20252084, in part by the National Key Research and Development Program of China under Grant 2021YFA1000500(4), in part by
	Shanghai Kewei under Grant 24DP1500500, 
	and in part by 111 Project under Grant B23008. \textit{(Corresponding author: Qiong Wu.)}}
\thanks{Bowen Tan and Qiong Wu are with the School of Internet of Things Engineering, Jiangnan University, Wuxi 214122, China, and also with the School of Information Engineering, Jiangxi Provincial Key Laboratory of Advanced Signal Processing and Intelligent Communications, Nanchang University, Nanchang 330031, China (e-mail: bowentan@stu.jiangnan.edu.cn, qiongwu@jiangnan.edu.cn).}
\thanks{Pingyi Fan is with the Department of Electronic Engineering, State Key Laboratory of Space Network and Communications, and the Beijing National Research Center for Information Science and Technology, Tsinghua University, Beijing 100084, China (e-mail: fpy@tsinghua.edu.cn).}
\thanks{Kezhi Wang is with the Department of Computer Science, Brunel University, London, Middlesex UB8 3PH, UK (e-mail: Kezhi.Wang@brunel.ac.uk).}
\thanks{Nan Cheng is with the State Key Laboratory of ISN and the School of Telecommunications Engineering, Xidian University, Xi'an 710071, China (e-mail: dr.nan.cheng@ieee.org).}
\thanks{Wen Chen is with the Department of Electronic Engineering, Shanghai Jiao Tong University, Shanghai 200240, China (e-mail: wenchen@sjtu.edu.cn).}
}
	\maketitle
	\begin{abstract}
			This letter proposes a novel three-tier content caching architecture for Vehicular Fog Caching (VFC)-assisted platoon, where the VFC is formed by
			the vehicles driving near the platoon. The system strategically coordinates storage across local platoon vehicles, dynamic VFC clusters, and cloud server (CS) to minimize content retrieval latency. To efficiently manage distributed storage, we integrate large language models (LLMs) for real-time and intelligent caching decisions. The proposed approach leverages LLMs' ability to process heterogeneous information, including user profiles, historical data, content characteristics, and dynamic system states.\color{black} Through a designed prompting framework encoding task objectives and caching constraints, the LLMs formulate caching as a decision-making task, and our hierarchical deterministic caching mapping strategy enables adaptive requests prediction and precise content placement across three tiers without frequent retraining.\color{black} Simulation results demonstrate the advantages of our proposed caching scheme.
	\end{abstract}
	
	
	\section{Introduction}
	\IEEEPARstart{I}{n} recent years, edge caching has been widely adopted to effectively reduce content transmission delay (CTD) for vehicular users requesting navigation, video, and entertainment services \cite{mobility}, \cite{Dong}, \cite{Di}. The integration of Internet of Vehicles (IoV) architecture and edge cloud infrastructure further accelerates the application of edge caching in reducing content retrieval latency \cite{IOV}. However, traditional content caching schemes mainly rely on fixed roadside units (RSUs), making it difficult to maintain stable connections. 
	
With the advancement of autonomous driving technology, vehicular platoons are widely studied for precise coordination and higher transportation efficiency, supported by efficient, stable and low-latency vehicle-to-vehicle (V2V) communications that reduce intra-platoon content retrieval delay \cite{sabu2024caching}, \cite{Zhou}, \cite{Yao}. We thus take vehicular platoons as the primary cache layer for latency reduction, and further leverage personal vehicles with local storage and stable platoon connections to form a vehicular fog caching (VFC) layer for assistance. This hierarchical structure underpins our proposed VFC-assisted platoon network.
	\begin{table}[!t]
		\centering
		\caption{\color{black}Performance Comparison of Different Caching Schemes}
		\label{tab:scheme_comparison}
        \color{black}	
		\begin{tabular}{|m{1.5cm}<{\raggedright}|m{1.4cm}<{\centering}|m{1.1cm}<{\centering}|m{1.1cm}<{\centering}|m{1.4cm}<{\centering}|}
			\hline
			& Model-Based Optimization & DRL Variants & Learning-to-Rank & LLM-Based Agents (Proposed) \\
			\hline
			Adaptability to Dynamics & Low & Moderate & Low & High \\
			\hline
			Latency & Low & Low & Low & Moderate \\
			\hline
			Retraining Cost & High & Very High & Moderate & Very Low \\
			\hline
			Context Awareness & Low & Moderate & Moderate & High \\
			\hline
			Scalability & Low & Moderate & High & High \\
			\hline
		\end{tabular}
	\end{table}

	\color{black}	
	As shown in Table \ref{tab:scheme_comparison}, traditional caching methods \cite{4}, \cite{Fan} such as model-based optimization, deep reinforcement learning (DRL) variants, and learning to rank (LTR) typically rely on precise scenario-specific modeling or large amounts of labeled data, which often lead to weak adaptability and considerable retraining costs in highly dynamic vehicular environments \cite{Wang}. In contrast, large language models (LLMs) \cite{5} leverage their rich prior knowledge and rapid comprehension of natural language instructions to adapt to dynamic conditions. Through prompt engineering, LLMs enable instant policy adjustment without extensive frequent retraining, making them more suitable for caching decisions in dynamic vehicular networks.
    \color{black}	
	
	The key contributions of this letter are as follows\footnote{The source code has been released at: 
		\href{https://github.com/qiongwu86/LLM-Empowered-Cooperative-Content-Caching-in-Vehicular-Fog-Caching-Assisted-Platoon-Networks}%
		{{https://github.com/qiongwu86/}\\
			{LLM-Empowered-Cooperative-Content-Caching-in-Vehicular-Fog-Caching-}
\\{Assisted-Platoon-Networks}}}

	\begin{enumerate}
		\item We propose an innovative VFC-assisted platoon caching architecture that utilizes platoon vehicles as local caching nodes and VFC vehicles as neighboring nodes. By placing content closer to the requester, this approach further reduces retrieval latency compared to traditional methods.
		\item We design a hierarchical prompt framework based on LLMs specifically for caching decisions, which utilizes multimodal data fusion to optimize content placement in a cooperative caching system and dynamically integrates real-time system data to enhance caching decisions.
    	\color{black}
		\item We propose a one-step LLM optimization mechanism without retraining, combined with a hierarchical deterministic caching mapping strategy to enable efficient dynamic adaptation, enhancing scalability and efficiency.
		\color{black}
			\vspace{-2.5em}
	\end{enumerate}
		\begin{figure}[!htbp] 
		\centering 
		\includegraphics[trim={2cm 5cm 0.9cm 1cm},clip,width=0.75\linewidth]{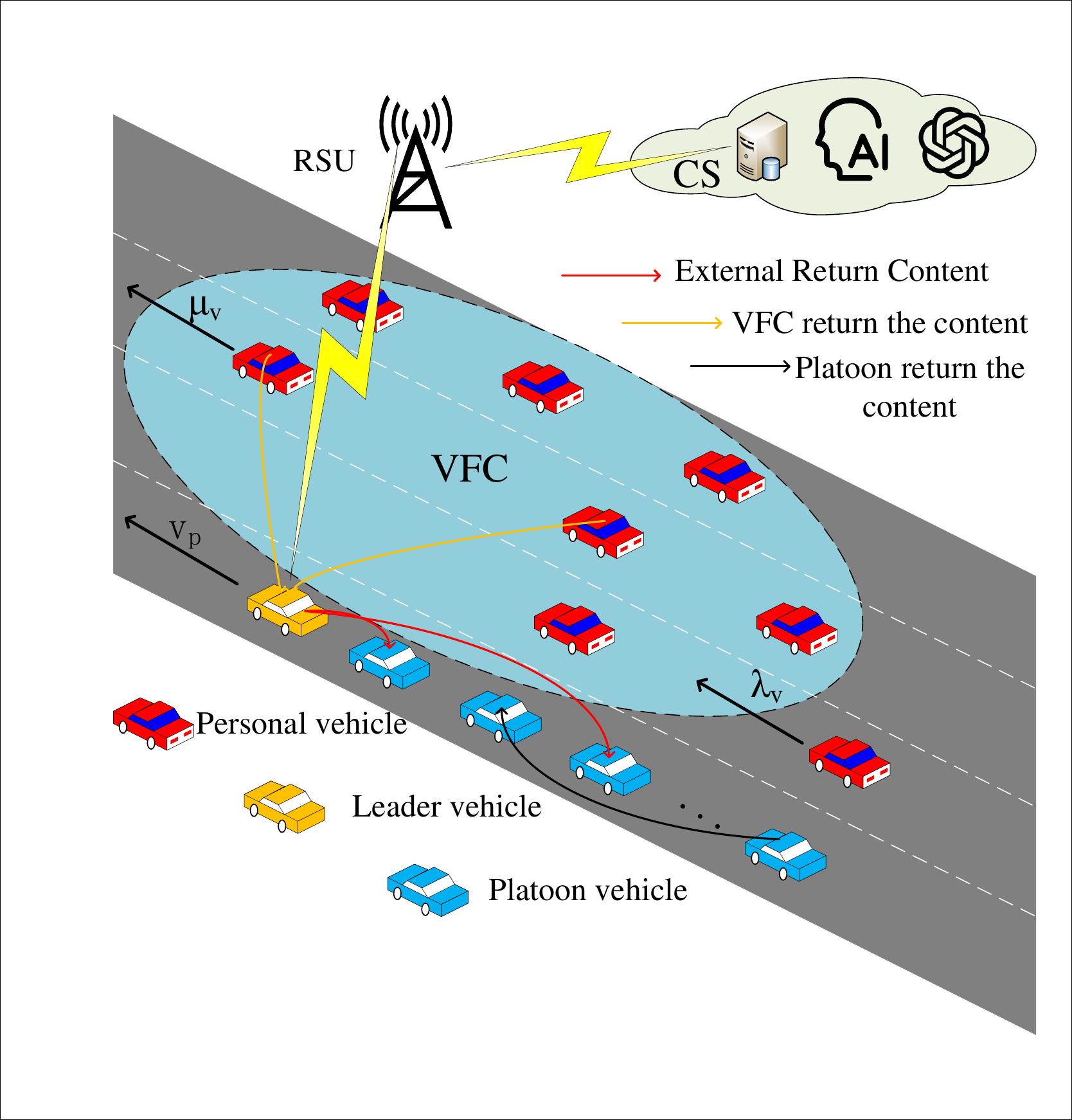} 
		\caption{System scenario} 
		\label{fig:visio} 
		\vspace{-1.5em}
	\end{figure}
	\section{System Model and Problem Formulation}
	\subsection{System Scenario}
	As shown in Fig.~\ref{fig:visio}, we consider a three-tier content caching system in an urban scenario, which consists of a platoon, a VFC comprising of personal vehicles within the communication range of the leader vehicle in the platoon, a local RSU connected to a cloud server (CS) and personal vehicles moving in and out the coverage of the VFC. The CS can provide all requested contents. We assume that platoon vehicles have the same capacity $M_p$. However, due to the heterogeneity of vehicles in the VFC \cite{Li2016Optimum}, each vehicle's capacity is different. We denote $\mathcal{M}^r = \{M_1^r,M_2^r,\ldots,M_k^r,\ldots,M_{N^r}^r\}$, where $N^r$ is the number of vehicles in the VFC in round $r$, $M_k^r$ is the capacity of $k$-th vehicle in the VFC. According to \cite {junyunfenbu}, we assume that $M_k^r$ follows a uniform distribution over the interval [$M_{\min}$, $M_{\max}$], which reflects the heterogeneity of different vehicle hardware configurations in a real vehicular networking environment.
	
	Each vehicle in the platoon stores extensive user historical data (ID, gender, age, occupation, content ratings), acting as both content provider and requester. Only platoon vehicles initiate requests and VFC vehicles merely provide cache. Thus, platoon vehicles could fetch content from the platoon, VFC, or RSU under different conditions. Specifically,

	 \textit {\textbf{1) Platoon}}: If a requested content is cached in the requesting vehicle itself, there is no retrieval delay. Otherwise, if it is cached in another vehicle within the platoon, the caching vehicle directly transmits the content to the requester.

	 \textit {\textbf{2) VFC}}: If a requested content is not cached in the platoon, the leader vehicle broadcasts the request to vehicles in the VFC. If the requested content is cached within the VFC network, the caching vehicle first transmits it to the leader vehicle. Subsequently, the leader vehicle forwards the content to the requesting vehicle. Notably, if the requesting vehicle is the leader vehicle, the forwarding phase is omitted.

	 \textit {\textbf{3) RSU}}: If a requested content is cached in neither the platoon nor the VFC, the leader vehicle relays the request to RSU. The RSU fetches the content from CS, and then sends it back to the requester in a manner similar to that of the VFC.
	\vspace{-1.5em}
	\subsection{Mobility and Communication Model}
	Consider a platoon of $N$ autonomous vehicles driving on a line with the same velocity $v_p$. For round $r$, the $i$-th vehicle in the platoon is denoted as $U_i^r$, thus the set of the platoon is denoted as $U = \{ U_1^r, U_2^r, \ldots, U_i^r,\dots, U_N^r \}$, where $U_1^r$ is the leader vehicle in the platoon. Considering that multiple users may be served by a single vehicle, we formally define the user subsets as $\mathcal{U}_i \subset \mathcal{U}$ for each vehicle in the platoon, where the complete user set satisfies $\mathcal{U} = \bigcup_{i=1}^{N} \mathcal{U}_i$ with $\mathcal{U}_i \cap \mathcal{U}_j = \emptyset$ for non-overlapping service scenarios. The total number of users is $N_u$. Throughout the entire rounds, we assume that the platoon is always within the communication range of the local RSU. Similarly, the set of VFC vehicles is denoted as $V^r = \{ V_1^r, V_2^r, \ldots, V_k^r, \dots,V_{N^r}^r\}$, where vehicles are sorted in ascending order of their distance to the leader vehicle:
	\begin{equation}
		\|V_1^r - U_1^r\| \leq \|V_2^r - U_1^r\| \leq \cdots \leq \|V_{N^r}^r - U_1^r\|,
	\end{equation}
	where, $\|\cdot\|$ denotes the Euclidean distance, and $V_k^r$ represents the $k$-th closest vehicle to $U_1^r$ in round $r$. $N^r$ may vary across rounds due to dynamic vehicle participation in the VFC. The arrival and departure rates of vehicles in the VFC are denoted as $\lambda_v$ and $\mu_v$. We assume that vehicles only enter or exit the VFC at intervals between each round. The maximum number of vehicles in the VFC is $K$ ($K \geq $1).  
	
	 In our system model, we employ Orthogonal Frequency Division Multiple Access (OFDMA) for both vehicle-to-infrastructure (V2I) and vehicle-to-vehicle (V2V) communications \cite{OFDMA}, \cite{Yang2012Doppler}. The model assumes sufficient Resource Block (RB) availability at the platoon, VFC nodes, and RSU to satisfy communication demands. Specifically, vehicles within the platoon can communicate using V2V links. According to the Shannon theorem, the transmission rate between $U_{i}^r$ and $U_{j}^r$ is calculated as
	 \begin{equation}
	 	R_{i,j}^{P,r}=B_{V2V}\log_2\left(1+\frac{P_{V}h_{i,j}^r}{\sigma_c^2}\right),
	 	\label{i2j}
	 \end{equation}
	 where $B_{V2V}$ is the V2V channel bandwidth, $h_{i,j}^r$ is the V2V channel gain for platoon vehicle $U_i^r$ and $U_j^r$, $P_{V}$ is the transmit power level used by vehicles, and $\sigma_c^2$ is the noise power \cite{Wu2015Performance}.
	 
	 Similarly, the transmission rate between the VFC vehicle $V_k^r$ and the leader vehicle $U_1^r$ in the platoon is calculated as
	 \begin{equation}
	 	R_{1,k}^{V,r}=B_{V2V}\log_2\left(1+\frac{P_{V}h_{1,k}^r}{\sigma_c^2}\right).
	 	\label{i2k}
	 \end{equation}
	 
	 Vehicle-to-RSU communication is possible through V2I links \cite{Wu2014Performance}, thus the transmission rate between the RSU and the leader vehicle $U_1^r$ is calculated as
	 \begin{equation}
	 	R_{1,R}^{r}=B_{V2I}\log_2\left(1+\frac{P_{R}h_R^r}{\sigma_c^2}\right).
	 	\label{v2r}
	 \end{equation}
	 
	\subsection{Problem Formulation}
	 Given the content set $\mathcal{F} \triangleq \{1, 2, \dots, f, \dots, N_c\}$, where $f$ is the $f$-th content item and each content $f$ has a uniform size $s$. For clarity, we assume that each content item represents a movie. Throughout this section, we use the index $i \in \{1, \dots, N\}$ to specifically denote the requesting vehicle $U_i^r$ in the platoon, and the index $j \in \{1, \dots, N\}$ to denote the caching vehicle $U_j^r$ in the platoon. 
	 
	 In round $r$, if content $f$ is cached in the platoon vehicle $U_j^r$, the content transmission delay is calculated as
	 \begin{equation}
		T_{i,j}^{P,r} = 
		\begin{cases} 
			 0 , & i = j, \\
			T_{i,j}^{P,r}=\dfrac{s}{R_{i,j}^{P,r}},  & i \neq j     ,               
		\end{cases}
		\label{eq:platoon_time} 
	\end{equation} 
	  where $i = j$ means the content is cached in the requesting vehicle, thus no retrieval delay, and $R_{i,j}^{P,r}$ is the transmission rate between $U_j^r$ and $U_i^r$ in the platoon, which has been expressed by \eqref{i2j}. If requested content $f$ is cached in vehicle $V_k^r$ in the VFC, the caching vehicle $V_k^r$ transmits the content to the leader vehicle $U_1^r$, which then forwards it to the requesting vehicle $U_i^r$, so content transmission delay is calculated as
	  \begin{equation}
	  	T_{i,k}^{V,r} = 
	  	\begin{cases} 
	  		\dfrac{s}{R_{1,k}^{V,r}} , & i = 1, \\[10pt] 
	  		\dfrac{s}{R_{1,k}^{V,r}}+ \dfrac{s}{R_{i,1}^{P,r}}, & i \neq1,                          
	  	\end{cases}
	  	\label{eq:trans_time} 
	  \end{equation}
	 where $R_{1,k}^{V,r}$ can be calculated by \eqref{i2k}, and $i = 1$ represents that the leader vehicle is the requesting vehicle. If content $f$ is neither cached in the platoon nor in the VFC, the RSU will fetch the content $f$ from CS, and sends the requested content to the leader vehicle $U_1^r$, which then forwards it to the requesting vehicle $U_i^r$, thus content transmission delay is calculated as
	 \begin{equation}
	 		T_{i,R}^{r} = 
	 		\begin{cases} 
	 				\dfrac{s}{R_{R-C}} +\dfrac{s}{R_{1,R}^{r}},         & i = 1,  \\[10pt]
	 			\dfrac{s}{R_{R-C}} + \dfrac{s}{R_{1,R}^{r}} + \dfrac{s}{R_{i,1}^{P,r}}, & i \neq 1,
	 		\end{cases}
	 		\label{eq:trans_time1} 
	 \end{equation}
	 where $R_{R-C}$ represents the backhaul link rate between RSU and CS, and $R_{1,R}^r$ has been calculated by \eqref{v2r}.
	 
	To efficiently represent caching decisions and better model vehicle’s retrieval latency across the platoon-VFC-CS tiers, we design a three-tiered decision structure as
	 \begin{equation}
	 	{X}^r = (x_{j,f}^r, y_{k,f}^r, z_f^r),
	 \end{equation}
	  \begin{align}
	 		x_{j,f}^r & \in \{0,1\}, j \in \{1, \dots, N\}, f \in \mathcal{F}, r \tag{Platoon Cache} \\
	 	y_{k,f}^r & \in \{0,1\}, \quad \forall k \in \{1,\dots,N^r\}, f \in \mathcal{F}, r \tag{VFC Cache} \\
	 	z_f^r     & \in \{0,1\}, \quad \forall f \in \mathcal{F}, r \tag{CS Fetch}
	 \end{align}
	 where $x_{j,f}^r=1$, $y_{k,f}^r=1$, and $z_f^r=1$ indicates content $f$ is cached in the platoon vehicle $U_j^r$, cached in the VFC vehicle $V_k^r$, and fetched from CS, respectively.

	 Each content can only reside in one tier to improve cache resource utilization. Therefore, the decision variables must satisfy the following constraint:
	 \begin{equation}
	 \sum_{j=1}^N x_{j,f}^r + \sum_{k=1}^{N^r} y_{k,f}^r + z_f^r = 1, \quad \forall f \in \{1,\ldots,N_c\}.
	\end{equation}
	  
	  For each round $r$, we define a binary request indicator variable $\alpha_{i,f}^r$, where $\alpha_{i,f}^r = 1$ indicates that vehicle $U_i^r$ requests content $f$ and $\alpha_{i,f}^r = 0$ indicates that $U_i^r$ does not request content $f$. Therefore, the delay for vehicle $U_i^r$ to fetch the content $f$, $T_{i,f}^r$, is given as
	\begin{equation}\hspace*{-0.4em}
		T_{i,f}^r = \alpha_{i,f}^r \left( \sum_{j=1}^{N} x_{j,f}^r T_{i,j}^{P,r} + \sum_{k=1}^{N^r} y_{k,f}^r T_{i,k}^{V,r} + z_f^r T_{i,R}^r \right),
		\label{iii}
	\end{equation}
	where $\alpha_{i,f}^r$ is generated by random sampling from the test set in each round. Furthermore, the VFC-assisted platoon content caching problem is equivalent to predicting content requests of vehicle $\alpha_{i,f}^r$  and determining an optimal caching decision ${X}^{r}$ to minimize content transmission delay: 
	 \begin{equation}
	 	\min_{{{X}^r}}  \frac{1}{N} \sum_{i=1}^{N} \sum_{f=1}^{N_c} T_{i,f}^r 
	 	\label{eq:objective_main}
	 \end{equation}
	 \begin{subequations}
	 	\renewcommand{\theequation}{\ref{eq:objective_main}\alph{equation}}
	 	\begin{align}
	 		&\text{s.t.} \ 
	 		&& x_{j,f}^r,\, y_{k,f}^r,\, z_f^r \in \{0,1\},
	 		&& \forall j, f, k, r,	 		
	 		\label{eq:var_def} \\%
	 		&&& \sum_{j=1}^N x_{j,f}^r + \sum_{k=1}^{N^r} y_{k,f}^r + z_f^r = 1,
	 		&& \forall f, r,
	 		\label{eq:exclusive} \\
	 		&&& \sum_{f=1}^{N_c} x_{j,f}^r = \frac{M_p}{s},
	 		&& \forall j, r,
	 		\label{eq:platoon} \\
	 		&&& \sum_{f=1}^{N_c} y_{k,f}^r = \frac{M_k^r}{s},
	 		&& \forall k, r,
	 		\label{eq:vfc}
	 	\end{align}
	 \end{subequations}

	 \begin{itemize}
	 	\item \eqref{eq:var_def}: All variables are binary, indicating decisions.
	 	\item \eqref{eq:exclusive}: The exclusive caching constraint ensures that each content resides in exactly one storage tier (platoon/VFC/CS) to prevent redundancy.
	 	\item \eqref{eq:platoon} and \eqref{eq:vfc}: The vehicle storage constraints ensure cached contents fully utilize the available capacity ($M_p$ for platoon vehicle and $M_k^r$ for VFC vehicle $V_k^r$) without exceeding it, thereby maximizing storage efficiency.
	 \end{itemize}

	\section{Proposed Approach}
	\subsection{Heterogeneous Information Modeling}
	\color{black}
Given LLMs' decision quality hinges on input prompts, we designed a concise systematic process to model and format multi-source heterogeneous data into structured text, enabling intelligent caching decisions. \color{black}At the beginning of round $r$, the framework collects heterogeneous information $\mathcal{H}^r$ from the current system scenario:
	\setcounter{equation}{11}
\begin{equation}
	\mathcal{H}^r = \Big\{ \mathcal{U}^r,\, \mathcal{R}^r,\, \mathcal{C},\, \mathcal{S}^r \Big\},
\end{equation}
\textbf{User profile}:
\begin{equation}\hspace*{-0.4em}
	\mathcal{U}^r = \Big\{ \big( \underbrace{\iota_\kappa^r}_{\text{ID}},
	\underbrace{g_\kappa^r}_{\text{Gender}},
	\underbrace{a_\kappa^r}_{\text{Age}},
	\underbrace{o_\kappa^r}_{\text{Occupation}} \big) \Big|
	\kappa \in \{1,\dots,N_u\} \Big\},
\end{equation}
\textbf{History Ratings}:
\begin{equation}\hspace*{-0.4em}
	\mathcal{R}^r = \big\{ (f, r_{\kappa, f})^r \mid f \in \{1,\ldots,N_c\}, \kappa \in \{1,\ldots,N_u\} \big\},
\end{equation}
\textbf{Content Type}:
\begin{equation}
	\mathcal{C} = \big\{ \big( f,\ \omega_f \big) \mid f \in \{1,\ldots,N_c\},\ \omega_f \in \mathcal{T} \big\},
\end{equation}
\textbf{System Cache State}:
\begin{equation}
	\mathcal{S}^r = \left(M_p \times N, \mathcal{M}^r \right),
\end{equation}
	\noindent where:
	\begin{itemize}
		\item $\iota_\kappa^r \in \mathbb{N}^+$: Unique user identifier in round $r$.
		\item $g_\kappa^r \in \{\mathrm{M}, \mathrm{F}\}$: Gender category (M = Male, F = Female).
		\item $a_\kappa^r \in \mathcal{A}$: Age cohort with predefined intervals:
	\begin{equation*}\noindent\hspace*{-1.5em}
		\mathcal{A} \!= \!\{\mathtt{Under\, 18}, \mathtt{18-24}, \mathtt{25-34}, \mathtt{35-45}, \mathtt{46-55}, \mathtt{56+}\}.
	\end{equation*}
		\item $o_\kappa^r \in \mathcal{O}$: Occupation taxonomy containing 21 categories:
		\begin{equation*}
			\mathcal{O} = \{\mathtt{Student}, \mathtt{Scientist}, \mathtt{Writer}, \ldots\}.
		\end{equation*}
		\item $(f, r_{\kappa, f})^r$ is user $\kappa$'s rating of content $f$ in round $r$.
		\item $(f,\ \omega_f \big)$ represents content $f$'s type is $\omega_f$,  $\omega_f \in \mathcal{T}$:
		\begin{equation*}
			\mathcal{T} = \{\mathtt{action}, \mathtt{drama}, \mathtt{comedy}, \mathtt{mystery}, \ldots\}.
		\end{equation*}
	\end{itemize}
		\subsection{Hierarchical Prompt Framework Design}
		Task description is crucial to LLMs for making inferential decisions \cite{YanLWC25}. We denote $D_\text{task}$ as the Language-based Task Description in order to help the LLM make more effective decisions, and the detailed description is as follows:
		\begin{taskdescription}
			\textbf{Task Goal:} Decide whether to cache movie $f$ and where to cache it to maximize hit ratio and minimize transmission delay.
			\textbf{Task Definition:} Select movies based on multi-dimensional information, without exceeding the cache capacity.
			\begin{itemize}[leftmargin=*,nosep]
				\item \textbf{System information description:}
				\begin{itemize}[leftmargin=0pt,nosep]
					\item Cache capacities: Platoon ($M_p\times N$), VFC ($ \mathcal{M}^r$).
					\item User profiles: $\mathcal{U}^r = \{[id, age, gender, occupation],\ldots\}$.
					\item History: $\mathcal{R}^r = \{[user\_id, movie\_id, rating], \ldots\}$.
					\item Movie type: $\mathcal{C} = \{[movie\_id , type], \ldots\}$.
				\end{itemize}
				\item \textbf{Output type:} A list of movie IDs, ${L}^r$,  representing the content placement sequence. Contents in ${L}^r$ are sequentially placed on the vehicles in the platoon, starting with the leader vehicle, until the cache capacities of all vehicles within the platoon are fully utilized. Remaining contents in ${L}^r$ are allocated to VFC vehicles in order of their increasing distance from the leader vehicle.
			\end{itemize}
			\textbf{Rules:}
			\begin{itemize}[leftmargin=*,nosep]
				\item Platoon consists of $N$ vehicles with the same cache capacity ($M_p$). VFC consists of personal vehicles within the communication range of the leader vehicle in the platoon.
				\item A total of $N_u$ users are evenly distributed across the vehicles in the platoon. Movies more likely to be accessed by each user should be placed closer to the user’s vehicle. The leader vehicle serves users with IDs 1 to $N_u/N$, the next vehicle serves IDs ${N_u/N+1}$ to $2N_u/N$, and so on.
				\item Strict priority: Platoon $>$ VFC (higher weights).
				\item Capacity limit: The number of movie IDs does not exceed the system cache capacity.
				\item No duplicate movie IDs in list ${L}^r$.
			\end{itemize}
		\end{taskdescription}
		\color{black}
		The selection of inputs in $\mathcal{H}^r$ and the optimized "constraint-reasoning-objective" prompt structure are designed to core caching requirements; in practical deployment, static content data are pre-stored to build a local knowledge base for controlling token length growth (with variations in $N_u$ and $N_c$), and lightweight LLMs are deployed on edge nodes to ensure inference speed meets real-time constraints.
		\color{black}
		
		To further enable LLMs to understand optimization problems and real-world scenarios, we design a novel hierarchical template structure to encode the input as
	\begin{equation}
		\mathcal{P}^r = {\text{Role}} \oplus {D_\text{task}}\oplus {\mathcal{H}^r},
	\end{equation}
	where $\mathcal{P}^r$ is the three-layer prompt we finally designed, which make LLMs more specialized in content caching in this environment. Each component serves distinct purposes:
	\begin{itemize}[leftmargin=*,nosep]
		\item \textbf{Role} establishes the LLM's decision-making role and knowledge boundaries. For example, the LLM defined as an edge caching expert would specialize in requests prediction and storage optimization, excluding non-caching domains.
		\item \textbf{Task description} $D_\text{task}$ encodes hard constraints including cache capacity and decision-making environment.
		\item \textbf{Heterogeneous information} $\mathcal{H}^r$ provide sufficient prior and current knowledge for LLM making decision.
	\end{itemize}
	\color{black}
		\subsection{Hierarchical Deterministic Caching Mapping Scheme}
	Our deterministic placement policy ensures feasibility via capacity constraints and non-duplication for the two-stage LLM-driven framework.  \color{black}This approach transforms a complex optimization problem into a two-stage process: an LLM generates a priority list, which is then executed by a deterministic policy, as defined follow:
	\begin{equation}
		{L}^r = \text{LLM}(\mathcal{P}^r),\quad
		{X}^r = \mathcal{G}({L}^r) .
		\label{eq:llm_mapping}
	\end{equation}
	
		\color{black}
	The LLM generates an ordered content list ${L}^r$, which is deterministically mapped to the caching decision ${X}^r$ through the policy of placement $\mathcal{G}$. Specifically, contents in ${L}^r$ are sequentially allocated to available cache capacities in platoon vehicles (following the predefined order in task description), and subsequently in VFC vehicles, until their capacities are filled. For any content $f \notin {L}^r$, the corresponding binary decision variables satisfy $x_{j,f}^r = 0$, $y_{k,f}^r = 0$ for all $j, k$, and $z_f^r = 1$. This approach is functionally equivalent to outputting only nonzero decision entries, thus reducing token overhead while maintaining functional equivalence.
	 \color{black}
	\section{Simulation Results}
	This section evaluates the proposed scheme's performance through a comparative analysis with several popular LLMs and baseline methods. We adopt the DRL\cite{L}, \cite{Li2019D2D} and the Python-based constraint programming solver CP-SAT as the methods for baseline experiments, where the DRL training parameters are set as $M$ = 15 and $K$ = 20. Some key parameters are listed in Table \ref{table:system_params}. We evaluate the proposed approach using two metrics, both averaged over $R$ simulation rounds: (1) Average Cache Hit Ratio (ACHR) and (2) Average Content Transmission Delay (ACTD), which respectively measure caching efficiency and content retrieval latency. ACHR is defined as the average probability of fetching requested content from vehicles in the platoon over $R$ rounds. If a requested content is cached in a platoon vehicle, it is referred to as a cache hit, otherwise, it is referred to as a cache miss. Thus, ACHR is calculated as
	
		\clearpage 
	\begin{figure*}[!t] 
		\vspace*{-2em} 
		\centering
		\begin{subfigure}[t]{0.3\textwidth} 
			\includegraphics[width=\linewidth]{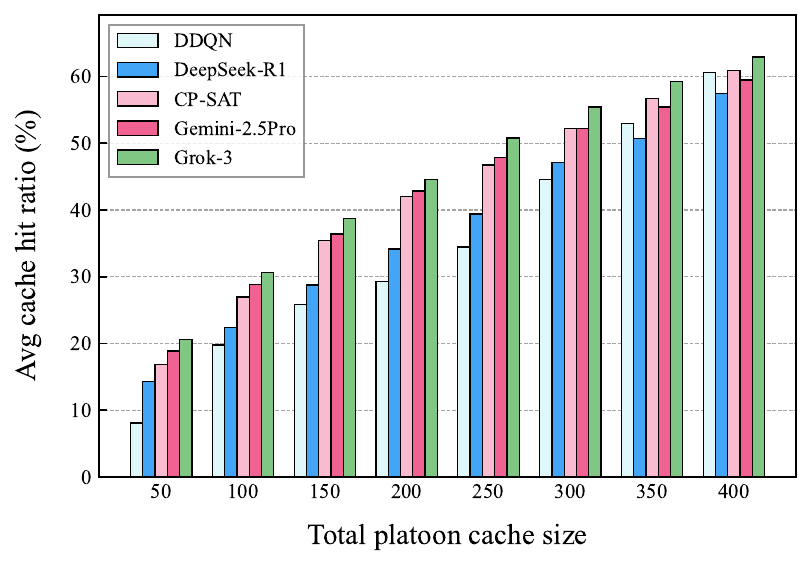} 
			\caption{\color{black}ACHR versus platoon cache size}
			\label{fig:bar_hit_ratio}
		\end{subfigure}%
		\hfill 
		\begin{subfigure}[t]{0.3\textwidth}
			\includegraphics[width=\linewidth]{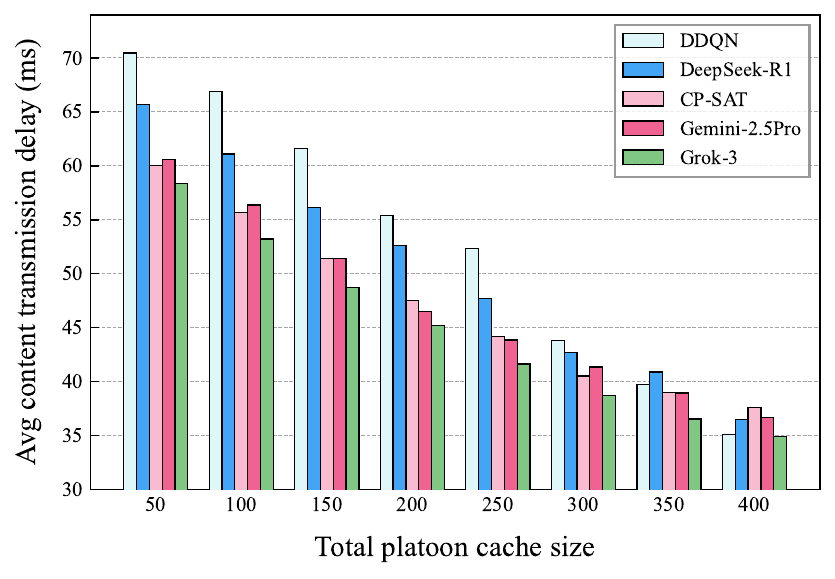}
			\caption{\color{black}ACTD versus platoon cache size}
			\label{fig:bar_latency}
		\end{subfigure}%
		\hfill
		\begin{subfigure}[t]{0.3\textwidth}
			\includegraphics[width=\linewidth]{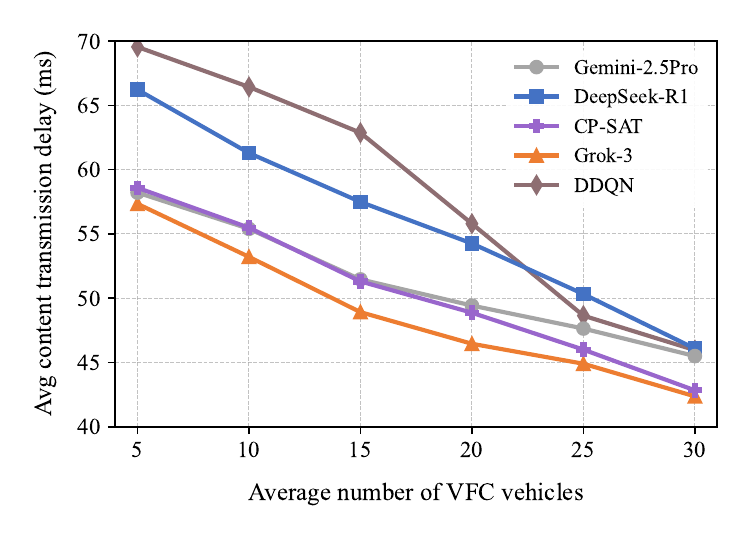}
			\caption{\color{black}ACTD vs. Avg. VFC vehicles}
			\label{fig:line_latency}
		\end{subfigure}
		\vspace{-0.5em} 
		\caption{\color{black}Performance comparison of caching schemes}
		\label{fig:combined_results}
	\end{figure*}

	\begin{equation}
		\mathrm{ACHR} = \frac{1}{R}\sum_{r=1}^R \frac{N_{\text{h}}^r}{N_{\text{h}}^r + N_{\text{m}}^r} \times 100\%,
	\end{equation}
	 where $N_{\text{h}}^r$ is the number of cache hits and $N_{\text{m}}^r$ is the number of cache misses in round $r$. ACTD indicates the average content transmission delay for all vehicles in the platoon over $R$ rounds, which is calculated as 
	\begin{equation}
	\mathrm{ACTD} = \frac{1}{R} \sum_{r=1}^R \frac{1}{N} \sum_{i=1}^{N} \sum_{f=1}^{N_c} T_{i,f}^r .
\end{equation}

	\begin{table}[htbp]
		\centering
		\caption{Parameters of System Model}
		\begin{tabular}{|c|c|c|c|}
		\hline
		\textbf{Parameter} & \textbf{Value} & \textbf{Parameter} & \textbf{Value} \\
		\hline
		$B_{V2I}$        & \SI{540}{\kilo\hertz}        & $B_{V2V}$                & 1 MHz     \\
		\hline
		$P_R$                & 30 dBm        & $P_V$              & 23 dBm         \\
		\hline
		$R_{R-C}$              & 0.8 Mbps       & $N$          & 10       \\
		\hline
		$v_p$   & 55 km/h        & $\sigma_c^2$        & -114 dBm         \\
		\hline
		$\lambda_v$           & 9        & $\mu_v$        & 8      \\
		\hline
		$s$           &100 bytes    &$M_{\max}$ &1500 bytes \\  
		\hline
		$M_{\min}$ &600 bytes   & $M_p$ &1000 bytes   \\
			\hline
		$R$ &12   & $N_u$ &30   \\
	
		\hline
			$\alpha$ &0.75   & $\beta$ &0.25   \\
		\hline
			$\gamma$ &0.002   & $N_c$ &2000   \\
		\hline
		\end{tabular}
		\label{table:system_params}
	\end{table}
	
	As illustrated in Fig.~\ref{fig:bar_hit_ratio}, we evaluate ACHR performance of five schemes (DDQN, CP-SAT, DeepSeek-R1, Gemini-2.5 Pro, and Grok-3) across different total platoon cache sizes ranging from 50 to 400 units, where $K=10$ and $s$ is one unit. The results show that all schemes demonstrate improved performance with increasing platoon cache capacity. Grok-3 consistently achieves the highest cache hit ratio across all cache sizes, outperforming other methods. \color{black}Its superior performance stems from the deep integration of retrieval-augmented generation (RAG) and task-adaptive reasoning, which mitigates domain knowledge gaps in caching decisions. Despite differing general reasoning rankings from Artificial Analysis \cite{artificial_analysis_2023}—Gemini-2.5 Pro (69/70), Grok-3 (67/70), DeepSeek-R1 (60/70)—Grok-3 outperforms others due to its more accurate capture of user-content correlations in platoon-VFC systems.\color{black}
	
	As shown in Fig.~\ref{fig:bar_latency}, ACTD of five schemes decreases with the increase of total platoon cache size. In general, Grok-3 outperforms the other solutions due to its excellent reasoning capacity and stability. However, LLMs’ performance gains diminish for larger cache sizes, limited by their output quality for voluminous content. \color{black} Moreover, DDQN achieves significant performance gains for cache sizes over 250, benefiting from its effective platoon-VFC cache balancing. \color{black}
	
	Fig.~\ref{fig:line_latency} illustrates that as the average number of vehicles in VFC increases, ACTD decreases accordingly, where $M_p=10$ (i.e., total platoon cache size is 100 units). The baseline DRL achieves the maximum reduction in ACTD when the average number of VFC vehicles reaches 20, as this matches its training environment. \color{black} By comparing Fig.~\ref{fig:bar_latency} and Fig.~\ref{fig:line_latency}, it is seen that enlarging the local platoon's cache space proves more effective at reducing Average Content Transmission Delay (ACTD) than deploying additional Vehicular Fog Computing (VFC) vehicles, as cached content serves as the primary mechanism for latency mitigation. \cite{Fan2010Study} \color{black}
	
	\section{Conclusion}
	\color{black}This work proposes an innovative LLM-Empowered three-tier cooperative caching architecture for VFC-assisted vehicular platoon, implementing dynamic content placement via a deterministic placement policy. \color{black} Our LLM-driven approach, integrating heterogeneous data via a prompt framework, enables adaptive decisions without the traditional need for extensive retraining. Although the LLMs perform well in this specific environment, their practicality under high-concurrency requests and generalization ability across multi-platoon scenarios still need to be tested. Future work includes multi-platoon collaboration, security, and energy-aware strategies.


\end{document}